\documentclass[reprint,twocolumn,superscriptaddress,showpacs,nofootinbib,notitlepage]{revtex4-1}
\usepackage[vcentermath]{}
\usepackage{epsf}
\usepackage{amscd}
\usepackage{amsmath}
\usepackage{amsfonts}
\usepackage{braket}
\usepackage{mathtools}
\usepackage{bm}
\usepackage{xcolor}
\usepackage[utf8]{inputenc}
\usepackage{graphicx}
\usepackage{latexsym,amsmath,amssymb,lmodern,float,url}
\usepackage{natbib}
\usepackage{color}
\usepackage{microtype}
\usepackage{slashed}
\usepackage{multirow}
\usepackage{comment}
\usepackage{bbm}
\usepackage{tikz}
\usepackage{graphics,setspace,epsfig,color}
\usepackage[colorlinks=true,backref=false, linktocpage=true,
citecolor=blue,urlcolor=blue,linkcolor=blue,pdfpagemode=UseOutlines]{hyperref}

\usepackage{qcircuit}

\hypersetup{%
  bookmarksnumbered=true,
  pdftitle = {},
  pdfsubject = {},
  pdfauthor = {},
  pdfkeywords = {}
}

\newcommand\tr{\mathop{\hbox{tr}}}

\newcommand{\beq}{\begin{equation}}
\newcommand{\eeq}{\end{equation}}
\newcommand{\bea}{\begin{eqnarray}}
\newcommand{\eea}{\end{eqnarray}}
\newcommand{\nn}{\nonumber}
\newcommand{\eq}[1]{Eq.~(\ref{eq:#1})}
\newcommand{\fig}[1]{Fig.~(\ref{fig:#1})}

\newcommand{\nb}{\mathbf{n}}
\newcommand{\pib}{\boldsymbol{\pi}}
\newcommand{\Jbb}{\mathbb{J}}
\newcommand{\Tbb}{\mathbb{T}}

\def\ket#1{\left| #1 \right\rangle}

\def\opbraket#1#2#3{ \left\langle #1 \left| #2 \right| #3 \right\rangle}

\def\pmat#1{\begin{pmatrix}#1\end{pmatrix}}

\begin{document}

\title[title]{ Sigma models on quantum computers}
\author{Andrei Alexandru}
\email{aalexan@gwu.edu}
\affiliation{Department of Physics, The George Washington University, Washington, D.C. 20052, USA}
\affiliation{Department of Physics, University of Maryland, College Park, MD 20742, USA}
\author{Paulo F. Bedaque}
\email{bedaque@umd.edu}
\affiliation{Department of Physics, University of Maryland, College Park, MD 20742, USA}
\author{Henry Lamm }
\email{hlamm@umd.edu}
\affiliation{Department of Physics, University of Maryland, College Park, MD 20742, USA}
\author{Scott Lawrence }
\email{srl@umd.edu}
\affiliation{Department of Physics, University of Maryland, College Park, MD 20742, USA}
\date{\today}
\collaboration{NuQS Collaboration}

\begin{abstract}
We formulate a discretization of sigma models suitable for simulation by quantum computers. Space is substituted by a lattice, as usually done in lattice field theory, while the target space (a sphere) is replaced by the ``fuzzy sphere", a construction well known from non-commutative geometry. Contrary to more naive discretizations of the sphere, in this construction the exact  $O(3)$ symmetry is maintained, which suggests that the discretized model is in the same universality class as the continuum model. That would allow for continuum results to be obtained for very rough discretizations of the target space as long as the space discretization is made fine enough. The cost of performing time-evolution, measured as the number of CNOT operations necessary, is $12 L T/\Delta t $, where $L$ is the number of spatial sites, $T$ the maximum time extent and $\Delta t$ the time spacing.
\end{abstract}

\maketitle

\section{Introduction}

The advent of quantum computers opens up a new method to attack several physics problems which have, up to now, remained intractable. Perhaps the most interesting of those is the numerical treatment of many-body/field theories theories with sign problems. In particular, the nonperturbative calculation of real time observables, where very little progress has been made up to now \cite{Berges:2006xc,Alexandru:2017lqr,Berges:2005yt}, is an obvious target for quantum computation. Of course, the hope of attacking these problems hinges on being able to formulate quantum field theories in a way suitable for quantum computers. This topic is still in its infancy. The naive expectation is that fermionic fields can be more easily implemented in quantum computers as a qubit can encode the presence or absence of a fermion in a given state. This is born out by the few existing calculations that have been performed on quantum comptuers~\cite{Martinez:2016yna,Klco:2018kyo,Lamm:2018siq}. Bosonic fields are not so simply implemented. The attempts made up to now involve either eliminating the bosonic fields using some special property of the model or truncating the occupation number at any given site~\cite{Hackett:2018cel,Macridin:2018gdw,Yeter-Aydeniz:2018mix,Klco:2018zqz,Bazavov:2015kka,Zhang:2018ufj,Unmuth-Yockey:2018xak,Unmuth-Yockey:2018ugm,Zache:2018jbt,Raychowdhury:2018osk,Kaplan:2018vnj,Stryker:2018efp}. The situation is analogous to the early days of (classical) computing in field theory. Classical bits also seem more amenable at describing fermionic than bosonic fields as the cost of storing and manipulating reasonable approximations to real numbers was too high to be practical in the early days.  There were at the time several attempts at substituting bosonic continuous field values by a finite set of values \cite{Lisboa:1982jj,Flyvbjerg:1984dj,Flyvbjerg:1984ji,Bhanot:1981xp,Jacobs:1980mk}.
In all  these schemes the symmetry of the model is reduced by the discretization of the bosonic fields.

When discretization reduces the symmetry it is unclear whether, in the spacetime continuum limit, the original model is recovered. For instance, the  non-linear sigma model in one spatial dimension with fields taking values on a sphere was studied in the approximation where the sphere is substituted by the vertices of a platonic solid. It seems to still be controversial whether the dodecahedron model is in the same universality class as the original spherical model \cite{Hasenfratz:2001iz,Caracciolo:2001jd,Hasenfratz:2000hd,PhysRevE.57.111,PhysRevE.94.022134,article}. In the case of gauge theories the question, at least for abelian theories, was settled long ago: abelian gauge theories with any finite discrete group $\mathbb{Z}_N$ are not in the same universality class as the $U(1)$ model and do not approach the $U(1)$ gauge theory as the spacetime continuum limit is taken \cite{Ukawa:1979yv,Creutz:1979zg,Grosse:1981bv}. 

This suggests that, to obtain the right continuum limit, we should construct a scheme where all the symmetries of the original model are maintained while discretizing the field variables in order to make the Hilbert space to have finite (and hopefully small) dimension. The topic of this paper is to present such a formulation. It is based on a well known construction in non-commutative geometry (the ``fuzzy sphere" \cite{Hoppe:1989aa,Madore:1991bw}) that has been used before in the study of (super)-membranes. The resulting system can be simulated on a quantum computer with two qubits per spatial site. We implement our simulation scheme on a simulated quantum computer and verify it produces the right results. The number of gates required is of the order of  $ \sim 12 L (T/\Delta t)$, where $L$ is the number of spatial sites, $T$ the maximum time extent and $\Delta t$ the time discretization step.

\section{Sigma model on the fuzzy sphere}
The $O(3)$ sigma-model is defined on a discretized space by the Hamiltonian
\bea\label{eq:H-sigma}
\mathcal{H} &=& 
\sum_r
 \left[ \frac{g^2}{2} \pib(r)^2 +\frac{1}{2g^2\Delta x^2} (\nb(r+1) - \nb(r))^2 \right ] \nn\\
 &=&\sum_r \left[ \frac{g^2}{2} \pib(r)^2 +\frac{1}{g^2\Delta x^2} (1-\nb(r+1)\cdot \nb(r)) \right] ,
\eea where $\nb$ is a unit three-dimensional vector, $\pib^2$ is the Laplace-Beltrami operator on $S^2$, and the sum runs over the $L$ spatial lattice sites. The global symmetry $\nb(r) \rightarrow O\cdot\nb(r)$, where $O$ is an orthogonal matrix, is evident. This model is asymptotically free in one spatial dimension \cite{1987gauge}. 

In the Hamiltonian formalism, the wave function is a function of $L$ copies of the sphere $S^2$, $\psi(\nb_1, \cdots, \nb_L)$. The Hilbert space is infinite dimensional even for $L=1$.
We will approximate this model by substituting the target space (the sphere) by the ``fuzzy sphere" \cite{Hoppe:1989aa,Madore:1991bw}. The fuzzy sphere is not defined as a subset of points of the sphere; instead, it is the functions on the sphere that are substituted by elements of a finite dimensional Hilbert space. 
Let us demonstrate the construction first in the $L=1$ case. The wave function of the system is a function of $\nb$ and can be expanded as
\beq\label{eq:psi_sphere}
\psi(\nb) = \psi_0 + \psi_i n_i + \frac{1}{2} \psi_{ij} n_i n_j + \hdots.
\eeq with the constraint $n_in_i=1$. In the fuzzy sphere regularization we substitute this Hilbert space by the space of matrices
\beq\label{eq:psi_fuzzy}
\Psi = \psi_0 \openone + \psi_i \Jbb_i + \frac{1}{2} \psi_{ij} \Jbb_i \Jbb_j + \hdots,
\eeq where $\Jbb_i$, $i=1,2,3$ are  generators of $SU(2)$  in a given representation $j$
(normalized such that $\sum_{i=1}^3\Jbb_i^2 =\openone$).
The important difference between \eq{psi_sphere} and \eq{psi_fuzzy} is that the sum in \eq{psi_fuzzy} terminates (for instance, if $j=1/2$, $\Jbb_i^2 \sim \sigma_i^2 = \openone$). Thus the infinite dimensional Hilbert space of functions on the sphere is substituted by a space of dimension $(2j+1)^2$. Since the space is finite dimensional it can be informally thought of as being the space of functions defined on a space with a finite number of points, the fuzzy sphere. In the $j=1/2$ case, the space of matrices $\Psi$ is four-dimensional and the fuzzy sphere can be informally thought of as a 4-point discretization of the sphere. However, the ``points" of the fuzzy sphere are ``spread out" and choosing them does not break rotation symmetry.
Notice that the fuzzy sphere is not defined as a subset of points of the sphere. It is the algebra of functions on the sphere that is deformed into a (non-commutative) algebra given by matrix multiplication.
Still, the fuzzy sphere is an approximation to the sphere in the sense that operation defined on the sphere can be approximated by equivalent constructions on the fuzzy sphere. For instance, the norm in the sphere and on the fuzzy sphere satisfy:
\beq
\frac{1}{2j+1} \tr (\Psi^\dagger \Psi) \xrightarrow[j\rightarrow\infty] {}\int_{S^2} \frac{d\Omega}{4\pi} |\psi|^2.
\eeq
 We refer to the references \cite{Madore:1991bw} for a discussion of the standard geometrical constructs of the sphere framed in terms of the algebra of functions (and their extensions to the fuzzy sphere). In particular, the Hamiltonian of the sigma model with one spatial site is simply the Laplacian on the sphere. (This describes the quantum mechanics of a free particle on the sphere.)
\beq\label{eq:kinetic}
-\frac{g^2}{2}\nabla^2 \psi \rightarrow H^0\Psi= \kappa\frac{g^2}{2} \sum_{k=1}^3[\Jbb_k, [\Jbb_k, \Psi]],
\eeq 
with $\kappa$ a normalization factor.  The eigenvalues of the Laplacian operator on the sphere are $l(l+1)$ 
for $l=0,1,\ldots$ with mulitplicities $2l+1$. When $\kappa=j(j+1)$ the spectrum of its fuzzy version $H^0$ is {\it exactly} the same but truncated to its lowest $(2j+1)^2$ values. This is in contrast to other discretizations of the sphere where the lowest eigenvalues are reproduced only {\it approximately}. Notice also that, as stressed before and contrary to other discretizations of the sphere, the fuzzy Laplacian $H^0$ has an exact $O(3)$ invariance $\Psi\rightarrow U(g)^\dagger \Psi U(g)$, where $U$ is the representation of the rotation $g$. 

From now on we will work with $j=1/2$ so the dimension of the Hilbert space at each site is 4. A convenient basis for this space
is: $\Tbb_0=i\openone/\sqrt{2}$ and $\Tbb_i = \sqrt{3/2}\ \Jbb_i$ which satisfies $\tr(\Tbb_a^\dagger \Tbb_b) = \delta_{ab}$.
In a system with $L>1$ spatial sites, the Hilbert space of the system is the tensor product of $L$ single-site Hilbert spaces. The generic wavefunction can be written as
\begin{align}
\Psi = &\sum_{a_0=0}^3\hdots \!\!\sum_{a_{L-1}=0}^3\psi_{a_{L-1}, \ldots, a_0} \ \ket{a_{L-1},\ldots,a_0} \,,\nonumber\\&\quad\text{with }
\ket{a_{L-1},\ldots,a_0} \equiv
\Tbb_{a_{L-1}} \otimes \cdots \otimes \Tbb_{a_0} \,.
\end{align}
The kinetic term ${\cal H}^0=\sum_n H^0(r)$ of the Hamiltonian is 
the sum of \eq{kinetic} acting on the Hilbert space of each site and it is diagonal in the basis 
$\ket{a_{L-1},\ldots,a_0}$ (for a single site operator $A$, 
$A(r)\equiv\openone^{\otimes L-r-1} \otimes A \otimes \openone^{\otimes r-1}$ denotes the operator acting on site $r$.)
In the $\Tbb$ basis the kinetic term is represented by a sum of similar tensor products of $4\times 4$ matrices with 
\begin{align}
h^0_{ij} &\equiv \opbraket{\Tbb_i}{H^0}{\Tbb_j} = \tr  \Tbb_i^\dagger H^0 \Tbb_j \nonumber\\&= \frac{\kappa g^2}2  \sum_{k=1}^3 \tr  \Tbb_i^\dagger [\Jbb_k,[\Jbb_k, \Tbb_j]]\,,\nonumber\\&\quad\text{and thus}\quad
h^0 = g^2\pmat{ 0 & 0 & 0 & 0\\ 0 & 1 & 0 & 0 \\ 0 & 0 & 1 & 0 \\ 0 & 0 & 0 & 1} \,.
\end{align}
The eigenvalues of ${\cal H}^0$ are then $E^0_{a_{L-1}...a_0}=g^2 \sum_{r} \chi({a_r})$, with $\chi(a)=1-\delta_{a0}$, that is, the kinetic energy of site $r$ is 0 if $a_r=0$ or $g^2$ otherwise.

%
%
%

The ``interaction" term arises from expanding the nearest-neighbor interaction term $-\nb(r+1)\cdot\nb(r)$ in \eq{H-sigma}: 
${\cal H}^I = \sum_r\sum_{k=1}^3 H^{Ik}(r+1,r)$ with $H^{Ik}(r+1,r) = -(\kappa/g^2\Delta x^2)(\Jbb_k)_{r+1} (\Jbb_k)_{r}$.
They involve only two neighbor sites at a time (one link). In the $\Tbb$ basis the $\Jbb_k$ operators are represented by the following
$4\times 4$ matrices $(j_k)_{ij}\equiv \opbraket{\Tbb_i}{\Jbb_k}{\Tbb_j}$:
\beq
j_1 = \openone\otimes\sigma_2/\sqrt3\,,\quad
j_2 = \sigma_2\otimes\sigma_3/\sqrt3\,,\quad
j_3 = \sigma_2\otimes\sigma_1/\sqrt3 \,,
\eeq
where $\openone$ is the two-dimensional identity matrix and $\sigma$'s are Pauli matrices.
The interaction term $H^{Ik}(r+1,r)$ in the $\Tbb$ basis is the matrix 
$h^{Ik}(r+1,r)=-({\kappa}/{g^2\Delta x^2}) (j_k)_{r+1} (j_k)_r$. By this we mean that the element
$\opbraket{a_{L-1},\ldots,a_0}{H^{Ik}}{a'_{L-1},\ldots,a'_0}$ is $h^{Ik}(r+1,r)_{i,j}$ where
$i$ and $j$ are the numbers that have the representation in basis 4 $a_{L-1},\ldots,a_0$
and $a'_{L-1},\ldots,a'_0$ respectively (the matrix indices run from $0$ to $4^L-1$.)


%

\section{Implementation of time evolution  and estimate of resources}

%

The  implementation of the time evolution of the model in terms of quantum gates starts by splitting the time evolution over a number of smaller steps $\Delta t=t/N$ and each time step using  the Suzuki-Trotter formula. Each time step is further split into the evolution due to the four parts of $H=H^0 + H^{I1}+ H^{I2}+ H^{I3}$  :
\beq\label{eq:trotter}
e^{-i H t} \approx
\left(
e^{-i H^{I3} \Delta t}
e^{-i H^{I2} \Delta t}
e^{-i H^{I1} \Delta t}
e^{-i H^{0} \Delta t}
\right)^N.
\eeq 
 The state of the system is time-evolved by first applying the kinetic term $e^{-i \Delta t \mathcal H^0}$ site-by-site; the site order does not matter since the $H^0(r)$ for different $r$ commute. We follow the kinetic term with the first interaction term $e^{-i \Delta t \mathcal H^{I1}}$. This evolution is done link-by-link, and again, the order does not matter, as all $H^{Ik}(r,r+1)$ commute with each other. We use periodic boundary conditions, so the evolution of the link from $r=L-1$ to $r=L$ is followed by evolution of a link from $r=L$ to $r=1$. The link-by-link evolution is repeated for $e^{-i \Delta t \mathcal H^{I2}}$ and $e^{-i \Delta t \mathcal H^{I3}}$. This completes one time step of the total time evolution. The whole process is then repeated $N$ times.

\begin{figure*}[!tbp]

\begin{equation*}
\scalebox{0.695}{%
    \Qcircuit @C=0.5em @R=0.5em @!R {
                \lstick{q_0} & \gate{\tt X} &\qw & \ctrl{1} &\qw & \gate{\tt X} & \qw\\
                \lstick{q_1} & \gate{\tt X} &\qw & \gate {\tt U_1(\Delta t)}\qw & \qw & \gate{\tt X} & \qw\\
                \lstick{q_2} & \gate{\tt X} &\qw & \ctrl{1} &\qw & \gate{\tt X} & \qw\\
                \lstick{q_3} & \gate{\tt X} &\qw & \gate {\tt U_1(\Delta t)}\qw & \qw & \gate{\tt X} & \qw\\
         }\kern 1em
     \Qcircuit @C=0.5em @R=0.5em @!R{
                \lstick{} & \gate{\tt S} & \gate{\tt H} & \ctrl{2} & \qw & \ctrl{2} & \gate{\tt H} & \gate{\tt S^\dag} & \qw\\
                \lstick{} & \qw & \qw & \qw & \qw & \qw & \qw & \qw & \qw\\
                \lstick{} & \gate{\tt S} & \gate{\tt H} & \targ & \gate{\tt U_1(-2\Delta t/3)} & \targ & \gate{\tt H} & \gate{\tt S^\dag}& \qw\\
                \lstick{} & \qw & \qw & \qw & \qw & \qw & \qw & \qw & \qw
                \gategroup{1}{7}{1}{8}{1.0em}{--}\gategroup{3}{1}{3}{3}{1.0em}{--}\\
         }\kern1em
  \Qcircuit @C=0.5em @R=0.5em @!R {
                \lstick{} & \qw & \qw & \ctrl{1} & \qw & \qw & \qw & \ctrl{1} & \qw & \qw& \qw\\
                \lstick{} & \gate{\tt S} & \gate{\tt H} & \targ & \ctrl{2} & \qw & \ctrl{2} & \targ & \gate{\tt H} & \gate{\tt S^\dag}&\qw \\
                \lstick{} & \qw & \qw & \ctrl{1} & \qw & \qw & \qw & \ctrl{1} & \qw & \qw & \qw \\
                \lstick{} & \gate{\tt S} & \gate{\tt H} & \targ  & \targ & \gate{\tt U_1(-2\Delta t/3)} & \targ & \targ & \gate{\tt H} & \gate{\tt S^\dag} & \qw \gategroup{1}{8}{2}{10}{1.0em}{--}\gategroup{3}{1}{4}{4}{1.0em}{--}\\
         }\kern1em
    \Qcircuit @C=0.5em @R=0.5em @!R {
                \lstick{} & \qw & \gate{\tt H} & \ctrl{1} & \qw & \qw & \qw & \ctrl{1} & \gate{\tt H} & \qw & \qw\\
                \lstick{} & \gate{\tt S} & \gate{\tt H} & \targ & \ctrl{2} & \qw & \ctrl{2} & \targ & \gate{\tt H} & \gate{\tt S^\dag} & \qw\\
                \lstick{} & \qw & \gate{\tt H} & \ctrl{1} & \qw & \qw & \qw & \ctrl{1}  & \gate{\tt H} & \qw & \qw\\
                \lstick{} & \gate{\tt S} & \gate{\tt H} & \targ & \targ & \gate{\tt U_1(-2\Delta t/3)} & \targ & \targ & \gate{\tt H} & \gate{\tt S^\dag} & \qw\gategroup{1}{8}{2}{10}{1.0em}{--}\gategroup{3}{1}{4}{4}{1.0em}{--}\\
         }}
\end{equation*}
      
    \noindent
    \caption{ Circuit implementing the time evolution. Starting from the left: the kinetic term $\exp[-i\Delta t{\cal H}^0]$ (for two sites), and the link terms: $\exp[-i\Delta t H^{I1}(1,0)]$, $\exp[-i\Delta t H^{I2}(1,0)]$, and $\exp[-i\Delta t H^{I3}(1,0)]$. 
    The notation for gates used here is standard in the quantum computing literature~\cite{nielsen2000quantum}.
    \label{fig:circuit}}
    \end{figure*}
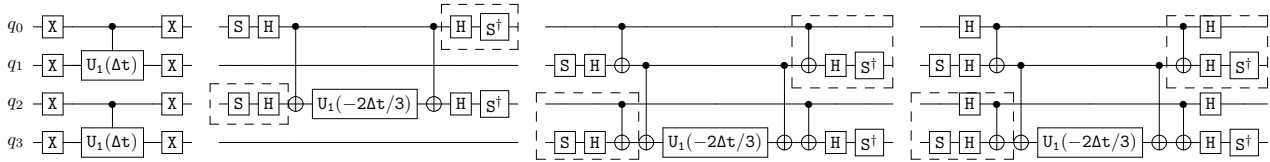

The local four-dimensional Hilbert space is encoded by two qubits so
\beq
\ket{a_{L-1},\ldots,a_0} \leftrightarrow 
|\underbrace{q_{2L-1}, q_{2L-2}}_\text{site\ L-1}, \cdots, \underbrace{q_1, q_0}_\text{site \ 0}\rangle
\eeq 
with the pair of qubits (for instance, $q_1 q_0$) being the binary digits of the value of the corresponding index $a$. For instance, $q_1=q_0=0$ corresponds to $a_0=0$ and $q_1=q_0=1$ corresponds to $a_0=3$.  In this basis, the kinetic term evolution at each site corresponds to the matrix
\begin{align}
e^{-i \Delta t h_0}=
e^{-i \Delta t }
\begin{pmatrix}
 e^{i \Delta t }& 0 & 0 & 0 \\
0 & 1& 0 & 0 \\
0 & 0 &  1 & 0 \\
0 & 0 & 0 & 1 \\
\end{pmatrix},
\end{align}
where from now on we set $g^2=1$ and $\kappa/g^2\Delta x^2=1$  for simplicity. The circuit implementing the kinetic term evolution is depicted on the left of \fig{circuit}.
Since ${\tt H S}^\dagger \sigma_2 {\tt S H} = \sigma_3$ the interacting term $H^{I1}=\frac 1 3 \openone\otimes\sigma_2\otimes\openone\otimes\sigma_2$ is related by a similarity transformation to $\frac 1 3 \openone\otimes\sigma_3\otimes\openone\otimes\sigma_3$, with the change of basis given by single qubit operations (here $h$ is the Hadamard and $S$ the phase one-qubit gates.)
Similarly, since ${\tt H} \sigma_1 {\tt H}  = \sigma_3$, 
$H^{I2}$ and $H^{I3}$ are related to $\frac 1 3 \sigma_3\otimes\sigma_3\otimes\sigma_3\otimes\sigma_3$, via  
similarity transformations involving only single qubit operations. 
We now use the fact that a controlled-NOT (CNOT)  gate implements a similarity transformation that takes $\sigma_3\otimes\sigma_3$ into $\sigma_3\otimes\openone$ and
that $\exp[i\theta \sigma_3\otimes\openone]$ is simply a rotation on the left qubit. 
For $H^{I1}$ we apply this for $q_2$ and $q_0$ qubit pair, whereas for the other two terms we have to apply the CNOT transformation
on the $(q_1,q_0)$ and $(q_3,q_1)$ pairs to reduce $\sigma_3\otimes\sigma_3\otimes\sigma_3\otimes\sigma_3$ to 
$\sigma_3\otimes\openone\otimes\sigma_3\otimes\openone$, and then a CNOT tranformation on the $(q_3,q_1)$ pair to reduce
the exponentiation to a single qubit rotation.
The quantum circuits implementing these operations are depicted in~\fig{circuit}.

%

\begin{figure}[!tbp]
      \centerline{\includegraphics[width=10cm]{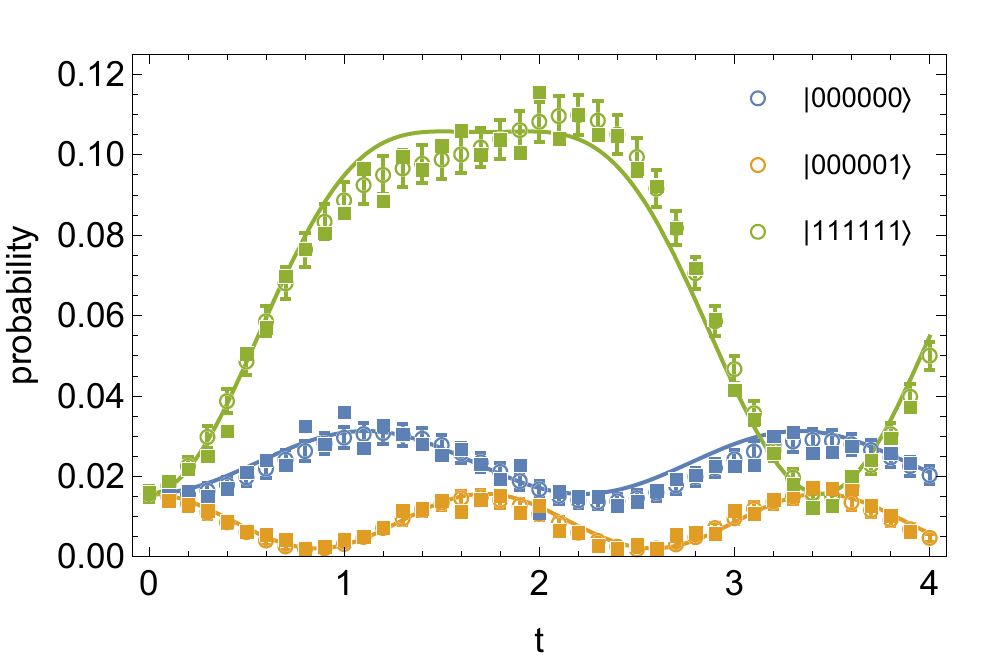}}
     \noindent
    \caption{Probabilities of states $\ket{000000}$, $\ket{000001}$, and $\ket{111111}$ starting from the initial state $(\ket{000000}+\ket{000001}+\cdots +\ket{111111})/\sqrt{64}$. The solid lines are the expected result obtained by direct diagonalization of the Hamiltonian. The empty circles are the results from the Trotter formula and their error bars the expected uncertainty following the binomial distribution. The filled points the result from the quantum simulator. Each data point from the simulator is the average of 4000 measurements.
   }
    \label{fig:simulation}
    \end{figure}

 The counting of quantum gates is, of course, dependent on the instruction set available on the hardware. However, the difficulty in the hardware implementation of two-qubit gates makes it unlikely that any two-qubit gate besides the CNOT gate will be available in hardware. CNOT gates are also, by far, the ones likely to generate decoherence. Thus, we will only count the number of CNOT gates required in our implementation.  
The kinetic term circuit has 2 CNOT gates per site (notice that a controlled-$U_1$ operation requires two CNOT gates to be implemented in usual architectures). The $\mathcal H^{I2}$ and $\mathcal H^{I3}$ link terms seems to require $6$ CNOT gates per site. However, since we apply $e^{i\delta t H^{I2}}$ on every link
the gates shown in the dashed box in \fig{circuit} cancel between
adjacent links and do not have to be applied. The result is that only $4$ CNOT gates per link are required.
A similar thing happens to the $\mathcal H^{I3}$ links. Finally, $\mathcal H^{I1}$ requires $2$ CNOT gates, for a total of 
 12 CNOT gates per site (for periodic boundary conditions, where there are as many links as sites). Since $T/\Delta t$ steps are needed for a total time evolution $T$, $36 (T/\Delta t)$ CNOT gates are required to implement $U(T)$ in our three-site model. This gate depth renders the model inaccessible to the current generation of processors. As a proof-of-principle, however, we run our algorithm in a quantum computer simulator (QISKIT~\cite{santos2017ibm,cross2017open})  for $L=3$ and the results are shown in~\fig{simulation}. The results in the figure were obtained with $\Delta t=0.2, \Delta x=1, g^2=1$ and show, for illustrative purposes, the probabilities of finding, as a function of time step,  the states $\ket{000000}, \ket{000001}$ and $\ket{111111}$ starting with the initial state with equal amplitudes of all elements of the basis $\ket{q_5 q_4 q_3 q_2 q_1 q_0}$ obtained by applying a Hadamard transformation on the state $\ket{000000}$. In the same figure we show the exact time evolution and ``Trotterized" evolution by multiplying the appropriate $4^3 \times 4^3$ matrices. The error bars reflect the expected variance from quantum mechanical measurement.

 It is important to stress that every step in our construction can be easily can be carried out in much bigger lattices and even in more spatial dimensions. The size of the blocks of time evolution -- involving at most 4 qubits at a time -- are independent of the system size. Also, the time evolution due to the kinetic term can be applied simultaneously to all sites and the evolution due to the hopping term can be applied simultaneously to half of the links at once. The method is essentially unchanged as the number of spatial dimensions is increased. Unfortunately, in the implementation in terms of quantum circuits we found, the number of CNOT gates is a little too large for current quantum computers available to us. Our attempts at running it on the IBM's {\tt ibmqx4} machine 
 resulted in mostly noise.

\section{Conclusions and Prospects}

One of the issues to be faced on the road to using quantum computers in quantum field theory is the presence of bosonic fields. 
The Hilbert space of a bosonic theory has an infinite number of dimensions {\it per spatial site}, while that for a fermionic theory is finite dimensional. On the other hand, the Hilbert space describing a quantum computer with a finite number of registers is finite dimensional. Thus, even after discretizing space, some further truncation of the field space will be required \cite{Hackett:2018cel,Klco:2018kyo}.  We propose a method to accomplish this while, at the same time, preserving the $O(3)$ symmetry of the  theory. 

There are two ways in which our fuzzy sphere model approximates the continuum $O(3)$ sigma model.  First of all, by increasing the dimension $2j+1$ of the representation of $O(3)$  the fuzzy sphere approaches the $O(3)$ sigma model defined by \eq{H-sigma}. Perhaps more interesting is the fact that the fuzzy model,  defined by $H=H^0+H^I$ (generalized to a large number of spatial sites), and the sigma model defined by the lattice hamiltonian \eq{H-sigma}, are likely to approach the same continuum limit as $\Delta t, \Delta x \rightarrow 0$. In fact, the continuum limit of the sigma model is obtained by tuning $\Delta t, \Delta x\rightarrow 0$ and $g^2$ is such a way as to keep physical quantities (mass gap, scattering amplitudes) fixed in physical units (perturbation theory indicates that the model is asymptotically free so the correct scaling is $g^2 \sim -1/\log(\Delta x)$~\cite{1987gauge}). In this limit, details of the Hamiltonian become irrelevant and any other Hamiltonian with the same field content and symmetries, on account of universality, give rise to the same continuum limit~\cite{zinn2002quantum}. More precisely, any other operators, consistent with the $O(3)$ symmetry, is of higher dimension and, presumably, irrelevant in the continuum limit. The reasonable assumption of universality can be checked in a classical calculation. The ``Trotterized" time evolution operator (\eq{trotter}) corresponds to an action discretized in both time and space. A Monte Carlo calculation using this action (analytically continued to  imaginary time) can demonstrate whether the fuzzy model is indeed in the same universality class as the sigma model and has, therefore,  the same $\Delta t, \Delta x \rightarrow 0$ limit.

Among theories of physical significance, bosonic fields also appear in principal chiral models (as, for instance, in low energy QCD) and gauge theories. These bosonic fields take values on group manifolds. A slight modification of the scheme proposed in the present paper can also be used in these cases, but it is somewhat more involved. A full account of these extensions will appear separately.

%



\begin{acknowledgments}
A.A. is supported in part by the National Science Foundation CAREER grant PHY-1151648 and by U.S. Department of Energy grant DE-FG02-95ER40907. A.A. gratefully acknowledges the hospitality of the Physics Departments at the Universities of Maryland where part of this work was carried out.
P.B., S.L., and H.L. are supported by the U.S. Department of Energy under Contract No.~DE-FG02-93ER-40762. The authors are grateful to Zohreh Davoudi for conversations on quantum computing, and to Jesse Stryker for helpful comments.
\end{acknowledgments}

\bibliographystyle{apsrev4-1}
\bibliography{discrete-groups,fuzzy-sphere} 

\end{document}